\documentclass[sigconf]{acmart}

\AtBeginDocument{%
  }
\usepackage{balance}

\usepackage{amsmath}
\usepackage{array}
\usepackage{geometry}
\usepackage{float}
\usepackage{xcolor}
\newcommand{\red}[1]{\textcolor{red}{#1}}

\begin{document}
\newcommand{\cmark}{\checkmark} 
\newcommand{\xmark}{\times}     

\title{Systematic comparison of deep generative models applied to multivariate financial time series.}

\author{Howard Caulfield}

\email{howard.cauflield@ul.ie}
\orcid{0009-0005-4706-7979}
\author{James P. Gleeson}
\affiliation{%
  \institution{MACSI, Department of Mathematics and Statistics}
  \state{University of Limerick}
  \city{Limerick}
  \country{Ireland}
}

\begin{abstract}
Financial time series (FTS) generation models are a core pillar to applications in finance. Risk management and portfolio optimization rely on realistic \emph{multivariate} price generation models. Accordingly, there is a strong modelling literature dating back to Bachelier's Theory of Speculation in 1901\cite{bachelier1901theorie}. Generating FTS using deep generative models (DGMs) is still in its infancy. In this work, we systematically compare DGMs against state-of-the-art parametric alternatives for multivariate FTS generation. We initially compare both DGMs and parametric models over increasingly complex synthetic datasets. The models are evaluated through distance measures over moment distributions for both the full and rolling FTS. We then apply the best performing DGM models to empirical data, demonstrating the benefit of DGMs on a implied volatility trading task.
\end{abstract}

\maketitle

\section{Introduction}
Generative methods in deep learning have launched overwhelming interest in the advent of artificial intelligence. However, the application of DGMs to time series is a burgeoning field. Historically, the field of time series modelling is dominated by econometric and mathematical modelling approaches. Given the success of deep learning in high dimensional fields, it offers an attractive approach to FTS applications. In section two we describe the current landscape of deep generative modelling with respect to FTS.
Section three focuses on the models we use and how we create and evaluate the synthetic datasets. 
We elucidate on our experimental implementation in section four. The results are reported and analysed in section five. Finally, in section six we conclude our analysis, outline caveats of our work and list further areas for research.
\section{Related Work}
The research landscape of synthetic time series generation and probabilistic forecasting using deep learning is quite fluid.
The general taxonomy of probabilistic generative models can be divided into implicit and explicit density modelling. Implicit models (General Adversarial Networks \cite{GAN}, Moment Matching Networks\cite{ADMM} and Diffusion models \cite{Diffusion}) do not require the user to imply a prior distribution in order for the data distribution to be learned. In contrast, explicit models (Variational Autoencoders \cite{VAE}, Normalizing Flows \cite{NORMALIZINGFLOWS}, Mixture Density Networks \cite{bishop1994mixture}, Boltzmann Machines \cite{RBM} and Autoregressive Density Estimation \cite{NADE}) do require an explicit prior, typically a Gaussian distribution. 

Early work on FTS using DGMs include FIN-GAN (\cite{FIN-GAN}) and Quant GAN (\cite{QUANTGAN}). FIN-GAN examined if GANs using multi-layer and convolution architectures could satisfy stylized facts of FTS (\cite{cont2001empirical}) such as volatility clustering. Quant GAN also examines this question using dilated convolutions (\cite{TCN}) in their architecture to capture long memory. TimeGAN \cite{Yoon2019TimeseriesGA} is a time series generation model which was tested over different time series tasks, including univariate stock generation. TimeGAN is comprised of both an autoencoder and GAN network. To induce temporal dynamics, they train a supervisor over the encoded latent space.

While financial price series are typically represented as financial returns other methods exist. Signatures \cite{lyons2014rough}, have been used to represent price paths. SigGWAN \cite{SIGWGAN} introduces a new measure Sig-W1 to compare time series models based on the Wassterstein distance of generated return signatures versus true return signatures. SigCWGAN \cite{SIGCWGAN} generates signatures of financial returns conditioned on a rolling window of signature features. The SigCGWAN model demonstrates superior Sig-W1 scores in comparison to TimeGAN, RCGAN \cite{RCGAN} and GMMN \cite{li2015generative}.

More recently, the application of DGMs in finance have been extended to specific applications. These include risk management applications such as tail-risk estimation (Tail-GAN \cite{TAILGAN}). ForGAN \cite{FORGAN} combines recurrent neural networks with GANs for probabilistic forecasts. Fin-GAN \cite{fin-gan1} extends ForGAN with an economic loss objective to improve portfolio Sharpe ratios. PAGAN \cite{mariani2019pagan} uses GANs to condition return generation on historical trends, and in doing so help guide portfolio optimization decisions.

Work that focuses on deep generative models for multivariate FTS deep generative models is limited. Hierarchical-SigCWGAN (H-SigCWGAN) is introduced in \cite{SIGHIGHDIM} and seeks to alleviate the dimensional bottleneck of signature approaches. The approach involves hierarchically clustering time series and determining a \emph{base} signature for each cluster. However, H-SigCWGAN does not demonstrate improvement in performance over its  counterpart SigCWGAN. CoMeTS-GAN \cite{Masi2023OnCS} builds on the work of \cite{QUANTGAN}. In CoMeTS-GAN, a correlation feature is also passed to the discriminator (lower triangular values of the correlation matrix). The model is trained on a Wasserstein loss \cite{wgan} and demonstrates improved correlated time series generation.
The authors of \cite{Tepelyan2023GenerativeML} use both variational autoencoders and normalizing flows to generate multivariate data for a basket of 500 stocks. The first part of the model comprises of a conditional importance weighted autoencoder model trained on predefined factors (PCA applied to a basket of indices). Using generated factor values, they learn a \emph{general} conditional normalizing flow model to generate multivariate time series. The performance of this two-step model is compared to parametric models (univariate GARCH and an exponential moving average model of the calculated PCA factors) and demonstrates superior negative log-likelihood values. 

The works most similar to our own include \cite{gatta2022neural}, \cite{Dogariu2022GenerationOR} and\cite{Ericson2024DeepGM}. In \cite{gatta2022neural}, multiple generative models are compared based on statistical properties, prediction scores and novelty. They focus on univariate price returns. The best performing model is an \emph{optimal} ensemble of SigCWGAN, TimeGAN, RCGAN and GMMN. However, how to create the optimal ensemble is not described. A large variation of DGMs (VAEs and GANs) is examined in \cite{Dogariu2022GenerationOR} with varying underlying architectures, primarily fully connected multi-layer or convolutional layer based networks. They train on empirical data with no conditioning. GANs are trained using a mixture of maximum mean discrepancy (MMD) and a standard GAN loss. After 100 epochs of training, variational autoencoders are found to work best. No measures of dependency (e.g., correlation) in the synthetic returns are reported . Lastly, \cite{Ericson2024DeepGM} is the closest to our work. Synthetic data is used to validate varying conditional generative models while comparing to historical simulation methods and parametric models. The work focuses on modelling value at risk of bond yields. A 250-day period is used for conditioning the models. Based on a mixed ranking of distribution distance, autocorrelation distance and a backtesting score (based on both training and testing data), the authors find that historical simulation outperforms both parametric and deep generative models. 

Our contribution is trifold. First, we introduce a systematic approach to test and compare multivariate price return generators. Through the systematic comparison of DGMs with incumbent state-of-the-art parametric methods, we show that DGMs can add value in multivariate financial return modelling. Lastly, we demonstrate and explore how conditional DGMs learn predictive features through a novel implied volatility trading task.

\section{Background}

\subsection{Models}
In the models below we assume $n$ instruments. Price returns $r_{t}$ refer to log returns at time $t$.

\subsubsection{Factor Stochastic Volatility}
The key assumption in factor volatility models is that there exists a vector of $m$ latent factors $f_t=(f_{1t},f_{2t},\cdots,f_{mt})$ which drive the $n$ observed returns $r_{t}=(r_{1t},r_{2t},\cdots,r_{nt})$ where $m<<n$. This factorization allows for specification of $m+n$ latent volatilities ($h_t$) which drive the system as opposed to $n \cdot \frac{n- 1}{2}$  unique entries of the full covariance matrix. The model (\cite{kastner2017efficient}) can be expressed as :
\begin{align}
    r_t&= \Lambda f_t +U_t(h_t^U)^\frac{1}{2} \epsilon_t, \\
    f_t&= V_t (h_t^V)^\frac{1}{2}\psi_{t},
\end{align}
where $\Lambda$ is the $n \times m$ factor loading matrix. The idiosyncratic latent variances is represented by $U_t(h_t^U)=diag(exp(h_{1t}),...,exp(h_{nt}))$, a diagonal $n \times n$ matrix, $V_t(h_t^V)=diag(exp(h_{(n+1,t)}),...,exp(h_{(n+m,t)}))$ is a diagonal $m \times m$ matrix that contains contains the factor variances. The variances are modelled as latent variables, the logarithm of variance follow an AR(1) process. Parameters are estimated using Markov chain Monte Carlo methods. For further details, see \cite{kastner2017efficient}.

\subsubsection{Multivariate GARCH (MGARCH)}
As the name implies, MGARCH is an extension to univariate GARCH models. The general GARCH model can be expressed as follows:
\begin{align}
   r_{t}|I_{t-1}&= \mu_{t}+\epsilon_{t},\\
   \epsilon_{t} &=H_t^{\frac{1}{2}} z_{t},
\end{align}
where $I_{t-1}$ represents the conditional information, e.g., previous returns. The ($n \times 1$) vector of price returns at time $t$ is represented by $r_t$, $\mu_t$ is the mean vector of returns with $\epsilon_t$ representing the innovation term. The residuals are modelled using $H_t$ which is the covariance matrix ($n \times n$) of the squared returns $r_t^2$ conditioned on previous squared returns and $z_t$ is an independent and identically distributed (i.i.d) random $n \times 1$ vector of mean 0 and standard deviation 1. The variance of returns follows:
\begin{align}
    V(r_t|I_{t-1})&=V_{t_1}(r_t)\\
    &=V_{t-1}(\epsilon_t)\\
    &=H_t^{\frac{1}{2}} V_{t-1}(z_t) {H_t^{\frac{1}{2}}}^{'}\\
    &=H_t,
\end{align}
where $V(r_t|I_{t-1})$ represents the conditional variance of returns. Key to different MGARCH implementations is the decomposition of $H_t$. For example, in the constant correlation model \cite{bollerslev1990modelling}, $H_t=D_{t}RD_{t}$ where $R$ represents the constant correlation amongst instruments and $D_{t}$ is a diagonal matrix ($n x n$)  of estimated volatility at time $t$.

In our experiments we use three models; the Dynamic Conditional Correlation model (DCC)\cite{engle2002dynamic},\cite{tse2002multivariate} with normal and Student-t distributed innovations and the Copula GARCH (COG) model \cite{jondeau2006copula}. Both the DCC and the COG models are implemented in the \emph{rmgarch} package \cite{ghalanos2022rmgarch}. Parameters are estimated using maximum likelihood. For further information on Multivariate GARCH models see \cite{bauwens2006multivariate},\cite{silvennoinen2009multivariate}.

\subsubsection{Deep Generative models}
We use a number of deep generative models for conditional price generation. The diversity of approaches is motivated by the question of whether explicit distribution modelling (e.g., normalizing flows ) vs. implicit distribution modelling (GAN based approaches) perform better at the task at hand. 

\begin{table}[ht]
    \centering
    \begin{tabular}{| m{2.5cm} | m{5cm} | m{7.5cm} |}
        \hline
        \textbf{Model Name} & \textbf{Architecture Description} \\
        \hline
        RCGAN \cite{RCGAN} & AR-FNN \cite{SIGCWGAN} 
        \\
        \hline
        TimeGAN\cite{Yoon2019TimeseriesGA} & Autoencoder, supervisor network over latent space and GAN 
        \\
        \hline
        GMMNs\cite{li2015generative} & AR-FNN \cite{SIGCWGAN}  
        \\
        \hline
        CoMeTS\cite{Masi2023OnCS} & Dilated Convolutional Layers in GAN structure 
        \\
        \hline
                CTVAE & VAE\cite{VAE} adopting AR-FNN structure
        \\
        \hline
                CTNF & Real NVP \cite{dinh2016density} with concatenating conditional vector.
        \\
        \hline
    \end{tabular}
    \caption{List of the DGM Models. We also considered a Wasserstein version of RCGAN with gradient penalty (RCWGAN). However this model performed poorly relative to RCGAN, so we do not include the results. We did not include SigCWGAN in the experiments due to the dimension requirements of the signature approach.}
    \label{tab:gan_models}
    \vspace{-.35cm}
\end{table}

We use the AR-FNN (Autoregressive feedforward neural network) architecture introduced in \cite{SIGCWGAN} as the base for a number of the models. AR-FNN works as follows, the output of one time step of the network is represented by $X_{t+1}=f(X_{t-wl:t},z_{t+1})$ where $f$ is the neural network, $X_{t-wl:t}$ represents a rolling window input of length $wl$ and $z_{t+1}$ represents the noise vector. $f$ typically includes residual blocks with parametric ReLU as an activation function. This format allows the model to generate time series of arbitrary length by iterative updating of the conditioning input value with the newly generated value i.e., $X_{t-wl:t}=cat(X_{(t-wl+1):t},X_{t+1})$.

\subsubsection{Heterogeneous Autoregressive model of Realized Volatility (HAR) \cite{corsi2009simple}}
To evaluate the empirical dataset, we use the HAR model for future realized volatility predictions.
The HAR model is defined as:
\begin{align}
\text{RV}_t^{HAR} = \omega + \beta_d \text{RV}_{t,d} + \beta_w \text{RV}_{t,w} + \beta_m \text{RV}_{t,m} + \varepsilon_t,
\end{align}
where $RV_t^{HAR}$ is the realized volatility at time $t$ predicted by the HAR model. The lagged daily, weekly, and monthly realized volatilities are given by $RV_{t,d}$,$RV_{t,w}$,$RV_{t,m}$ respectively and $\omega, \beta_d, \beta_w, \beta_m$ are the fitted intercept and daily, weekly and monthly realized volatility coefficients. The error term $\varepsilon_t$ is assumed to be i.i.d with mean zero and variance $\sigma^2$.
\subsection{Datasets}
The datasets used can be divided into synthetic and empirical. The synthetic datasets were developed with increasing complexity, see table \ref{tab:features_comparison}. NGARCH and Heston were chosen as the base generative models of the more complicated datasets. Both of these models can be parameterized to satisfy many of the stylized facts observed in FTS. Given the parametric multivariate models are derived from the same family of the synthetic datasets, this should provide a challenging synthetic environment to test the relative ability of DGMs. To extend Heston to the multivariate setting we used the approach outlined in \cite{dimitroff2011parsimonious}.

The NGARCH(1,1) model can be defined as 
\begin{align}
r_t &= \mu + \varepsilon_t, \\
\varepsilon_t &= \sigma_t z_t, \\
\sigma_t^2 &= \omega + \beta (\varepsilon_{t-1} - \gamma \sigma_{t-1})^2 + \alpha \sigma_{t-1}^2,
\end{align}
where $r_t$ is price return, $\mu$ is the mean return, $\varepsilon_t$ is the error term, $\sigma_t$ is the conditional standard deviation, and $z_t$ is an i.i.d. standard normal random variable. The parameters of the model are $\omega, \beta, \gamma$ and $\alpha$. The introduction of the $\gamma$ parameter (in comparison to the standard GARCH model) adjusts variance for return innovations, introducing a leverage effect. If $\gamma$ is positive, this will reduce the impact of positive innovations i.e., $\varepsilon_{t-1}>0$ and equally exacerbate negative innovations.

The Heston model is described by the following stochastic differential equations:
\begin{align}
    dS_t &= \mu S_t \, dt + \sqrt{V_t} S_t \, dW_t^S, \\
    dV_t &= \kappa (\theta - V_t) \, dt + \sigma \sqrt{V_t} \, dW_t^V,
\end{align}
where $S_t$ is the asset price at time $t$, $\mu$ is the drift rate of the asset price, $V_t$ is the variance of the asset price at time $t$, $\theta$ is the long-term variance, $\kappa$ is the rate at which $V_t$ reverts to the long-term variance $\theta$, $W_t^S$ and $W_t^V$ are two Brownian motions with correlation $\rho$ and $\sigma$ is the volatility of the variance process.

The Heston+ and GARCH+ datasets include regime components. After generating the time series for a burn-in period, we allow  the correlation structure to vary using predetermined correlation matrices based on a rolling volatility measure over all $n$ instruments versus a volatility percentile level i.e., when the average volatility of the basket of instruments increases above a certain level, we change the correlation structure of the time series.

For Heston+, we model the jump component with a Poisson distribution. Jump rates and sizes depend on a cyclical regime. The jump regimes (normal and large) are based on a cyclical probability level which aims to \emph{naively} replicate stock earning seasons. The large jump regime is guaranteed to happen at least once semi-annually i.e, 126 trading days.

\begin{table}[ht]
    \centering
    \begin{tabular}{| m{1.5cm} | m{0.75cm} | m{0.75cm} | m{0.95cm} | m{0.75cm} | m{0.75cm} | m{0.75cm} | m{0.75cm} |}
        \hline
        \textbf{Features} & \textbf{Corr} & \textbf{AR} & \textbf{ARCH} & \textbf{CBM} & \textbf{Reg} & \textbf{Jump} \\
        \hline
        NGARCH & $\cmark$ & $\cmark$ & $\cmark$ & $\cmark$ & $\xmark$ & $\xmark$ \\
        \hline
        Heston & $\cmark$ & $\cmark$ & $\cmark$ & $\cmark$  & $\xmark$ & $\xmark$ \\
        \hline
        NGARCH+ & $\cmark$ & $\cmark$ & $\cmark$ & $\cmark$ & $\cmark$ & $\xmark$ \\
        \hline
        Heston+ & $\cmark$ & $\cmark$ & $\cmark$ & $\cmark$ & $\cmark$ & $\cmark$ \\
        \hline
    \end{tabular}
    \caption{We tested the models across increasingly challenging and more realistic datasets. AR (adds path dependency), ARCH (adds an autoregressive nature to the variance of the series), CBM (introduces multi-modal dependencies via a correlation block model), Reg (introduces volatility regimes into the data) and Jumps (adds discrete jump events based on regime factors).}
    \label{tab:features_comparison}
    \vspace{-.8cm}

\end{table}
The empirical dataset comprises of daily bid and ask of the close prices for equity options and the closing prices of the underlying stock for 50 instruments (all components of the S\&P500). The dataset ranges from February 2010 until April 2024.  All the underlying price data is transformed to log returns and adjusted for stock splits and dividends. 
We build a straddle (call and put combined) dataset from the option data. The strike closest to the expiry forward is found. From this, we approximate the long (short) straddle price  as the straddle bid (ask) plus (minus) three quarters of the straddle bid-ask spread. We estimate realized profit (gamma profit and theta loss) from the next day underlying moves using standard straddle Greek approximation formulas \cite{sinclair2013volatility}. 

\subsection{Evaluation Measures}

To evaluate the models with the synthetic data we looked primarily at distribution distance measures over the entire generated returns and also rolling windows of the generated returns. The rolling window is one third of the full time series length.
The  we examine are mean, standard deviation, skew and kurtosis. We also look at the distribution of correlation values.

We report the Earth Mover's Distance (EMD \cite{bonneel2011displacement}) of the true and generated moment distributions. This distance represents the minimum required work required to transform distribution $P$ to $Q$ over a distance measure. We use the squared euclidean for the distance measure. Finally, we rank each method across all measures and report a naive equal weighted rank for the most complex datasets  in table \ref{tab:ranks}. 

To determine the quality of the models in an economical application with the empirical data, we create a \emph{factor-like} volatility basket. The basket is based on option theta neutrality (i.e., the derivative of option value with respect to time). For the top and bottom $n$ stock expected realized volatility (from the HAR model) to implied volatility ratios, we invest in an equal weighted theta neutral basket of total value long/short one theta. For example, with $n=5$, we invest 0.2 theta in each of the 5 top ranked ratios by buying the straddle and equivalently we sell 0.2 worth of theta for the 5 lowest ranked ratios. The profit/loss (PnL) of this basket determines the strength of the model.

Lastly, given the complexity of the empirical dataset, we examine how well the models learn dynamic correlation in an empirical setting. We inspect the differences of the averaged daily Jaccard Index between past, future and generated correlation networks.
The correlation networks are created using bootstrapped samples, similar to the work of \cite{wang2023topological}. 

\begin{table}[ht]
    \centering
    \begin{tabular}{|l|c|}
      \hline
         Abbreviation & Model   \\
        \hline
        FSV & Factor Stochastic Volatility \\
        FSV\textsubscript{R} & Rolling FSV \\
        FSV\textsubscript{C} & Regime based FSV \\
        DCCN &  DCC-GARCH with normal innovations  \\
        DCCN\textsubscript{R} & Rolling DCCN \\
        DCCT & DCC-GARCH with Student-t innovations \\
        DCCT\textsubscript{R} & Rolling DCCT \\
        COG & Copula GARCH \\
        COG\textsubscript{R} & Rolling COG \\
        \hline
    \end{tabular}
    \caption{Model Abbreviations}
    \label{tab:ModelAbbreviations}
        \vspace{-.75cm}

\end{table}

\section{Experiments}

We compare all models over all synthetic datasets, after which we select the best performing DGMs to test on the implied volatility trading task.

\subsection{Implementation Notes}
Much of the implementation extends previous work (\cite{SIGWGAN}, \cite{Yoon2019TimeseriesGA}, \cite{VAE}, \cite{dinh2016density} and \cite{Masi2023OnCS}), with minor adjustments. Similarly, the FSV and MGARCH models used rpackages \emph{factorstochvol} \cite{fsvpackage} and \emph{rmgarch} \cite{ghalanos2022rmgarch} respectively. We use a mixture of python and R (leveraging the rpy2 package) for the implementation.
All input was raw log returns, we did not find any benefit in return normalization.  
We run all experiments over 5 random seeds. To ensure that the generated time series for training, validation and testing is not stationary, we join together varying parameterizations of each model consecutively, i.e., for a dataset of length 25000 we may include 50 different parameterizations of length 500. We divide all synthetic datasets into 60\% training, 20\% validation and 20\% testing. We base the hyperparameter selection on a preliminary grid search over all DGM models parameters for the validation testing, with early stopping based on distribution distance measures. The empirical dataset is split into  60\% training, 10\% validation and 30\% testing sets. 
Across all conditioned models we use a conditioning matrix of size (50$\times$40) i.e., 40 time steps for 50 instruments. We evaluate the quality of generated samples by comparing the distribution of generated samples to the distribution (and distribution of time series properties) of the next true 40 time steps for all 50 instruments.

For the rolling FSV models, we used the parameters of full model as priors for each rolling model. Furthermore, as recommended by \cite{kastner2017efficient} we restrict the factor loadings matrix to upper triangular. We use a burn-in of 500 samples and 5000 draws with thinning set to every fifth draw.

The copula for the Copula GARCH model uses Kendall correlation with a multivariate Student-t distribution for the copula form.

\begin{table*}[ht]
    \centering
    {\fontsize{8}{12}\selectfont
    \begin{tabular}{|l|c|c|c|c|c|c|c|c|c|c|c|c|c|c|c|}
        \hline
        Measure & CoMeTS & CTNF & CTVAE & GMMN & RCGAN & TimeGAN & FSV\textsubscript{C} & COG & COG\textsubscript{R} & DCCN & DCCN\textsubscript{R} & DCCT & DCCT\textsubscript{R} & FSV & FSV\textsubscript{R} \\
        \hline
        Corr & 1.83 & 2.96 & 2.19 & 0.06 & \red{0.02} & 2.76 & 2.07 & \red{0.02} & 2.99 & \red{0.02} & 1.38 & 0.03 & 3.23 & 2.10 & 2.21 \\
        Kurt & 0.25 & 1.01 & 0.21 & 0.02 & \red{0.01} & 0.20 & 0.06 & 0.09 & 0.08 & 0.02 & 0.39 & 0.05 & 0.68 & 0.21 & 0.27 \\
        Mean & 0.33 & 0.05 & 0.35 & 0.09 & 0.07 & \red{0.03} & \red{0.03} & 0.04 & 0.10 & 0.12 & \red{0.03} & \red{0.03} & 0.08 & 0.04 & \red{0.03} \\
        Skew & 0.13 & 0.22 & 0.18 & 0.01 & \red{0.00} & 0.02 & 0.03 & 0.03 & 0.08 & 0.02 & 0.20 & 0.02 & 0.26 & 0.06 & 0.09 \\
        Std & 2.36 & 0.52 & 27.06 & 0.14 & \red{0.03} & 0.45 & 0.08 & 0.10 & 0.42 & 0.04 & 7.83 & 0.16 & 0.93 & 0.62 & 0.05 \\
\hline
        Corr\textsuperscript{R} & 4.24 & 31.49 & 27.65 & 2.29 & \red{0.03} & 18.92 & 17.39 & 0.05 & 25.27 & 0.08 & 16.29 & 0.08 & 9.13 & 17.40 & 14.22 \\
        Kurt\textsuperscript{R} & 0.06 & 2.62 & 0.31 & 0.07 & 0.12 & 0.38 & 0.22 & 0.08 & 0.27 & 0.08 & 0.29 & \red{0.05} & 0.93 & 0.22 & 0.12 \\
        Mean\textsuperscript{R} & 0.47 & 0.63 & 2.54 & 0.20 & \red{0.06} & 0.64 & 1.34 & \red{0.06} & 0.37 & 0.09 & 0.25 & 0.07 & 0.19 & 0.87 & 0.17 \\
        Skew\textsuperscript{R} & 0.07 & 0.69 & 0.24 & 0.39 & 0.13 & 0.49 & 0.13 & \red{0.04} & 0.21 & 0.07 & 0.14 & 0.08 & 0.35 & 0.15 & 0.06 \\
        Std\textsuperscript{R} & 1.02 & 2.26 & 18.69 & 0.49 & \red{0.04} & 4.41 & 17.26 & 0.21 & 0.66 & 0.20 & 4.00 & 0.34 & 2.15 & 11.49 & 0.46 \\
        \hline
    \end{tabular}
    \caption{EMD distance measures for NGARCH+ Dataset, red font indicates the lowest (i.e., the best) respective score. The $R$ superscript in measures identifies the rolling distribution values}
    \label{tab:NGARCH}}
    
\end{table*}

\begin{table*}[ht]
\centering
\vspace{-15pt}
{\fontsize{8}{12}\selectfont
\begin{tabular}{|l|c|c|c|c|c|c|c|c|c|c|c|c|c|c|c|}
\hline
Measure & CoMeTS & CTNF & CTVAE & GMMN & RCGAN  & TimeGAN & FSV\textsubscript{C} & COG & COG\textsubscript{R} & DCCN &  DCCN\textsubscript{R} &  DCCT & DCCT\textsubscript{R} & FSV & FSV\textsubscript{R} \\
\hline
Corr & 6.29 & 1.29 & 0.32 & 2.83 & \red{0.09}  & 4.44 & 0.25 & 0.94 & 9.05 & 1.28 & 3.97 & 0.17 & 4.69 & 0.60 & 0.21 \\
Kurt & 24.91 & 22.61 & 19.55 & 5.52 & 3.73 & 22.97 & 4.78 & 24.59 & 17.28 & 23.00 & 10.92 & 24.29 & 9.16 & \red{3.48} & 3.72 \\
Mean & 0.47 & 0.05 & 0.65 & \red{0.02} & 0.18  & 0.10 & 0.05 & 0.04 & 0.04 & 0.05 & 0.40 & 0.03 & 0.06 & 0.20 & 0.03 \\
Skew & 6.70 & 6.00 & 5.00 & 1.30 & 0.84  & 6.06 & 1.10 & 6.46 & 4.52 & 6.07 & 2.48 & 6.37 & 2.05 & \red{0.79} & 0.88 \\
Std & 0.82 & 3.34 & 10.67 & 0.08 & 2.25  & 1.17 & 0.14 & 2.22 & 0.12 & 3.15 & 2.82 & 0.17 & 0.21 & 0.71 & \red{0.04} \\
\hline
Corr\textsuperscript{R} & 10.40 & 10.29 & 11.64 & 14.65 & 9.04  & 9.22 & 9.03 & 7.43 & 13.59 & 10.90 & \red{3.46} & 9.41 & 4.18 & 11.57 & 7.73 \\
Kurt\textsuperscript{R} & 19.53 & 18.20 & 9.33 & 3.77 & 2.74  & 6.79 & 3.61 & 20.10 & \red{2.38} & 8.20 & 3.40 & 14.63 & 3.69 & 3.45 & 4.22 \\
Mean\textsuperscript{R} & 1.40 & 0.49 & 2.19 & 0.19 & 0.55  & 0.57 & 0.17 & 0.28 & 0.21 & 0.31 & 0.30 & \red{0.16} & 0.24 & 0.40 & 0.24 \\
Skew\textsuperscript{R} & 6.00 & 4.36 & 2.39 & 1.18 & 1.14  & 1.91 & 0.94 & 4.92 & 0.67 & 2.02 & \red{0.82} & 3.55 & \red{0.82} & 0.90 & 1.05 \\
Std\textsuperscript{R} & 6.99 & 6.83 & 10.63 & \red{0.21} & 2.23  & 2.85 & 0.43 & 4.60 & 0.38 & 4.29 & 1.57 & 1.89 & 0.57 & 1.67 & 0.82 \\
\hline
\end{tabular}
\caption{EMD distance measures for Heston+ Dataset.}
\label{tab:HESTON}}
\end{table*}

\begin{table}[ht]
    \centering
    \vspace{-15pt}
    \begin{tabular}{|l|c|c|c|}
        \hline
        & NGARCH+ & Heston+ & Combined \\
        \hline
        RCGAN & \red{3.0} & 6.6 & \red{4.80} \\
        FSV\textsubscript{R} & 6.6 & \red{4.3} & 5.45 \\
        DCCT & 3.8 & 7.9 & 5.85 \\
        GMMN & 6.0 & 5.9 & 5.95 \\
        FSV\textsubscript{C} & 8.1 & 4.4 & 6.25 \\
        COG & 3.9 & 10.9 & 7.40 \\
        DCCN & 4.1 & 11.0 & 7.55 \\
        COG\textsubscript{R} & 9.8 & 6.3 & 8.05 \\
        FSV & 9.9 & 6.7 & 8.30 \\
        DCCN\textsubscript{R} & 9.7 & 7.3 & 8.50 \\
        DCCT\textsubscript{R} & 11.4 & 6.2 & 8.80 \\
        TimeGAN & 9.7 & 10.5 & 10.10 \\
        CoMeTS & 8.6 & 14.3 & 11.45 \\
        CTNF & 12.6 & 11.6 & 12.10 \\
        CTVAE & 12.8 & 12.4 & 12.60 \\
        \hline
    \end{tabular}
    \caption{Average algorithm combined score ranking. Columns are sorted based on ascending combined rank. The combined rank is the averaged rank performance on both datasets. The lower the rank, the better the performance. }
    \label{tab:ranks}
        \vspace{-1.15cm}

\end{table}

\subsection{Model Adjustments}
\subsubsection{DGM}
In general for the models we found that including absolute price returns in addition to price returns improved performance. This could be considered analogous to learning over a drift and scale component. For the GMMN implementation, inspired by CoMeTS-GAN \cite{Masi2023OnCS}, we altered the loss function to include MMD losses over the absolute returns and also the lower triangle correlation values. Similar to other implementations we learn the GMMN loss over multiples of a base bandwidth. We approximate the base bandwidth length using the median of pairwise squared euclidean distances. We found that misspecified bandwidth choice can have a significant effect on learning capability. 
\subsubsection{Parametric Models}
To allow a fair comparison to DGM approaches, we needed to introduce some conditionality to the parametric models. The base parameteric models (i.e., no subscript in the model abbreviation) are trained over the entire training dataset. We also include rolling window versions for both parametric methods, where we train models over window length 40 (similar to the conditioning vector of DGMs).
The rolling approach for FSV parameter estimation did not always provide valid covariance matrices for generation. We only include results for models that did so.
Lastly, as parameteriszation of FSV models typically require larger training sets than rolling windows of length 40, we created a regime based approach for FSV. We use functional clustering to create a number of models based on the previous 40 timesteps. We first approximate the cumulative return of the past 40 timesteps for each instrument using polynomial splines of order six. We normalize across degree order and cluster curve types using Gaussian Mixture Models (GMM).
We then represent each cumulative return curve by its cluster.
This reduces the conditional vector to $50 \times 1$. We then cluster these representations, based on their groupings, again using GMM. We define these clusters as regimes. For each regime, we stitched together return series of the training dataset and learned a FSV model for each. For testing, we identified regimes through the two-step functional clustering approach and generated time series accordingly. To determine the number of factors for each FSV model, we used scree plots. The parametric model abbreviations are listed in table \ref{tab:ModelAbbreviations}.

\subsection{Empirical Application}

For the empirical application, we include the following restrictions. By default we trade the straddle of the nearest expiry, however if the stock has an expected earnings event within five days we trade the second nearest expiry. We do this to prevent the model from trading event volatility e.g., earnings events typically lead to contango in the volatility term structure. By trading the implied volatility of the further expiry, we limit this effect. Lastly, the straddle realized profit approximations are dependent on the strike being close to the forward price. We rule out any trades where the straddle strike is more than 50 basis points from the forward price.
To derive features from the generated data, we take the expected future daily, weekly and monthly realized volatility over all generated batches.
We substitute these features into the baseline HAR model. We also extend the HAR model by including some network-based realized volatility features. We define the additional feature as a degree weighted summation of realized volatility per instrument. The neighbours are defined based on the conditioning vector correlation matrix i.e., if correlation is greater than correlation threshold we insert a link into the adjacency matrix. The motivation for these additional features is to determine if the generated data is maintaining informative relationships between instruments.
Similar to the work of \cite{clements2021practical}, we estimate the HAR model with ridge regression and an exponential weighting scheme.

\begin{figure}[htbp]
    \centering
        \centering
        \includegraphics[width=.45\textwidth]{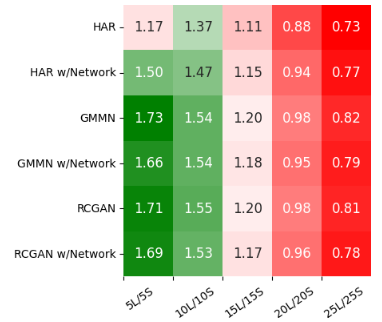}
        \caption{PnL from Long/Short Basket. This table represents the profit per day from trading the long/short volatility basket (higher numbers are better). On the y-axis the model type is listed with or without network based features. The x-axis describes the content of the basket, going from stronger signals on the left (top 5 vs. bottom 5 ranked predicted realized vs. implied volatility ratios) to all signals being traded in the last column i.e., equal-weighted basket of 25 long vs. 25 short.}
        \label{fig:subfiga}
        \end{figure}

\begin{figure}[htbp]
\centering        \includegraphics[width=.45\textwidth]{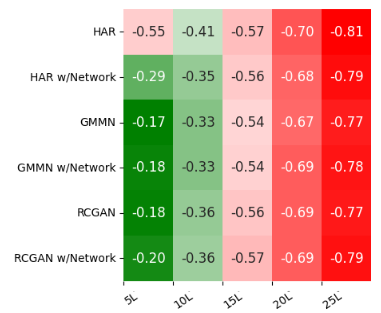}
        \caption{Similar to figure \ref{fig:subfiga}, this table shows the PnL from long-only side of the signal.}
        \label{fig:subfigb}

\end{figure}
\begin{figure}[htbp]
\centering
        \includegraphics[width=.45\textwidth]{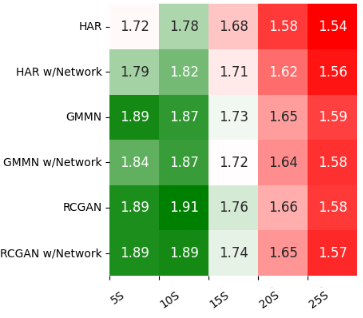}
        \caption{Similar to figure \ref{fig:subfiga}, this table shows the PnL from short-only side of the signal.}
        \label{fig:subfigc}
    \label{fig:PNL}
\end{figure}
\begin{figure}[htbp]  
  \centering
  \includegraphics[width=6.5cm,height=4.5cm]{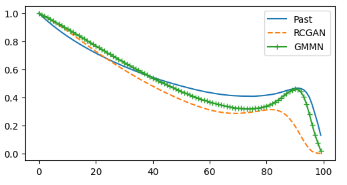}  
  \caption{\small The y-axis represents the average Jaccard index of the future correlation network to the past and model generated correlation networks for all time steps in the test dataset. The x-axis represents the percentile for the correlation threshold i.e., 90 represents a correlation network based on the $90^{th}$ largest percentile correlation value for a sliding window of length 40. A value of one implies perfect match. The blue line (past correlations) falls away from one quite dramatically indicating how dynamic correlation values are in the empirical dataset. In the higher percentile regions, we observe that both generated models underperform versus the past correlation network. This implies that the generated network features used in the HAR model are poorly formed, shedding light on the slight degradation in performance of the generated HAR model with network features.
  }
  \label{fig:aji}  
  \vspace{-.35cm}
\end{figure}

\section{Results}
\subsection{Synthetic Datasets}
We only report the results for the most complicated synthetic datasets, NGARCH+ and Heston+. The increased complexity of these datasets highlight more clearly the differences in model performance.  Results are averaged over five random seeds.
Tables \ref{tab:NGARCH} and \ref{tab:HESTON} detail the earth movers distances for all models over the NGARCH+ and Heston+  datasets respectively. 
Table \ref{tab:ranks} summarizes the average rank of each model per dataset across all distance measures. 
For the NGARCH+ dataset, RCGAN is the clear best performer, however unsurprisingly the GARCH specified models perform well. 
For the Heston+ dataset, there is no consistent superior model. We find that the FSV models score relatively well, with the rolling model the best relative performer. This however comes with the caveat that we only report results with a valid estimated covariance matrix. GMMN is the best performing DGM model for Heston+ with RCGAN a close second. The parametric models performed well in relation to their specified datasets. Yet the performance of both RCGAN and GMMN is quite promising. The general scores in Heston+ highlight the increased difficulty of this dataset. In summary, RCGAN performs relatively best out of all models, ranking high consistently across all categories. 

\subsection{Empirical Dataset}
We report the realized profit per day (PnL) (excluding vega profit and transaction fees) in figures \ref{fig:subfiga}, \ref{fig:subfigb} and \ref{fig:subfigc} for the long/short, long-only and short-only baskets respectively. Signal strength across all strategy combinations is clear i.e., the more \emph{select} the signal e.g., top 5 vs. bottom 5 in comparison to top 25 vs. bottom 25 provides a monotonically improving result. The HAR model with network features outperforms HAR by itself, this echoes work done in \cite{chen2022multivariate}.  The difference between the generative based HAR and the baseline is stark, with clear outperformance using the generative data. To investigate this further we examine PnL of both the short and long baskets. The generative models perform better in both baskets but when comparing the top 5/bottom 5 baskets, the majority of increased performance comes from the long side  (-0.55 to -0.18) vs. (1.72 to 1.89). There is little to no difference in the performance of the generative models. Notably the network features add no value to the generative HAR models (see figure \ref{fig:aji} for possible reasons). The baseline HAR model is the only model which does not have a monotonically increasing performance with basket composition. We examined the range of PnL per basket constituents (i.e, max PnL over all instruments minus min PnL over all instruments). The baseline model has the largest PnL range in the 5L/5S basket, nearly twice that of it's corresponding 10L/10S basket confirming that the variance spikes for the baseline model in this signal bucket. The PnL range for all other models across baskets is much more stable. This implies the trading signal from the other generative HAR models is more robust relative to the baseline with respect to dispersion of PnL across instruments.
 
We show the average Jaccard index for different thresholds in figure \ref{fig:aji} for window length 40. The relationship exhibited is similar for both models, however the performance of GMMN degrades less. Despite learning correlation in the synthetic datasets, neither model manages to identify conditional empirical drivers of correlation dynamics.  

\begin{figure}[ht]  
  \centering
  \vspace*{-.3cm}
  \includegraphics[width=.5\textwidth]{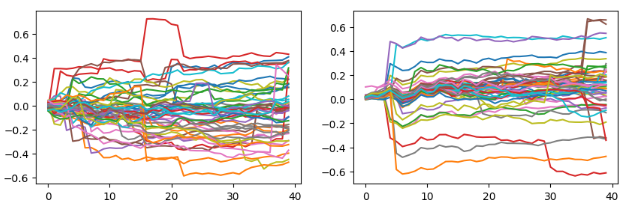}  

  \caption{\small Above is a sampled and true version of the Heston+ dataset. While capturing jumps, there are some clear differences to the true dataset. RCGAN exhibits greater constant lower volatility within the dataset and jumps are not as closely clustered together like the true sample.}
  \label{fig:jumps}  
   \vspace{-.8cm}
\end{figure}

\subsection{Further results for RCGAN}
\begin{figure}[ht]  
  \centering
  \includegraphics[width=.5\textwidth]{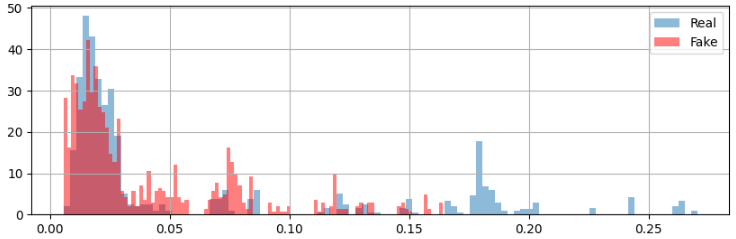} 
  \caption{\small While successfully learning the majority of return characteristics some artifacts still exist. In this histogram, we see the frequencies of rolling standard deviation values for each time series generated. The best model has trouble learning the bi-modality (second blue peak around 0.17) of the rolling standard deviation.}
  \label{fig:std}  
      \vspace{-.35cm}

\end{figure}

\begin{figure}[ht]  
  \centering
  \includegraphics[width=.5\textwidth]{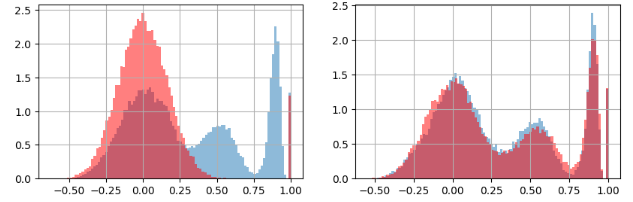}  
  \caption{\small The histograms represent the normalized count of correlation values for the synthetic (red) and true (blue) data. The figure on the left shows the correlation structure of the synthetic returns early in the training phase. After training the  figure on the right shows that RCGAN manages to learn the correlation structure.}
  \label{fig:cbm}  
      \vspace{-.35cm}

\end{figure}

Given the strong performance of RCGAN, it warranted further exploration.  While the model can generate jumps (figure \ref{fig:jumps}), one particular difficulty for the model was trying to capture rolling bimodality of the standard deviation introduced by regimes (figure \ref{fig:std}).
However, it is also quite capable of capturing multimodal relationships, capturing correlation block model specifications as seen in figure \ref{fig:cbm}.

\section{Conclusion \& Discussion}

In this work, we have introduced a systematic framework for analysing multivariate financial price series models. To the best of our knowledge, this is the first comparison of DGMs to state-of-the-art multivariate parametric models with challenging synthetic datasets. After demonstrating impressive performance on synthetic datasets relative to parametric models, we highlight the additional value gained by DGM models in a novel implied volatility trading task. 

Despite extensive efforts, we could not improve the \emph{relative} performance of the other DGM approaches to RCGAN and GMMN within the same training time. However, given the success of these models in other fields, it is likely there is room for improvement. With this in mind, the ease of training both RCGAN and GMMN warrants merit. 
Both implicit distribution models, RCGAN and GMMN perform  quite well to all models which require explicit distribution priors. Examination of the returns generated by the other GAN implementation (CoMeTS) found very smooth price generations, potentially due to the dilated convolution operations over shorter conditioning time frames (in comparison to the original implementation). The lower number of assumptions for DGMs also offer an advantage over FSV models. For FSV we firstly used scree plots to find a suitable number of factors to describe the data and had to rely on correctly specified covariance matrices for data generation.

There are numerous potential extensions to this work. The additional benefit of generations from both GMMN and RCGAN models in improving the HAR model performance implies that generative price return models could act as \emph{foundation} models from which we can further build economic applications, in addition to directly learning models with applications (\cite{fin-gan1},\cite{TAILGAN}). With a \emph{foundation} model in mind, we limited our work to 50 instruments, increasing the number of instruments may lead to improvement of the model (as demonstrated by \cite{Tepelyan2023GenerativeML}). The benefit of increased data is also motivated by the concept of universal price features (\cite{sirignano2021universal}). Determining how the models studied here perform with a greater number of instruments is an open question.

As with any experiment, our approach is not exhaustive and required many design choices. Due to limitations of time we did not include diffusion models which have shown state-of-the-art performance in other fields \cite{songscore} but doing so is a straightforward extension. Additionally, the performance degradation from including jumps based on an exogenous factor (the cyclical probability to mimic earnings season) raises the question of whether a more informative conditioning vector could improve results e.g., returns and implied volatility.

While we demonstrated the importance of network effects in the HAR model baseline, as evidenced by the empirical correlation network experiment, the models do not learn dynamic correlation or leverage the natural network representation as in \cite{wang2022network}. The use of GNNs are prominent in FTS forecasting problems (\cite{Xiang2022TemporalAH}, \cite{Cheng2022FinancialTS}, \cite{Chen2018IncorporatingCR}) but to the best of our knowledge there currently exists no deep graph-based financial return generators. This avenue of research offers a natural way to learn and generate dynamic multivariate price return relationships.

\begin{acks}
  This work was partly funded by Science Foundation Ireland under grant numbers 18/CRT/6049 (H.C.) and 12/RC/2289 P2 (J.P.G.).
\end{acks}

\vspace{-8pt}
\bibliographystyle{ACM-Reference-Format}
\bibliography{references}

\balance
\end{document}